\documentclass[prd,aps,a4paper,superscriptaddress,twocolumn,nofootinbib]{revtex4}
\usepackage{graphicx}
\usepackage{color}
\usepackage{xcolor}
\usepackage{dcolumn}
\usepackage{bm}
\usepackage{slashed}
\usepackage{amsmath}
\usepackage{latexsym}
\usepackage{amssymb}
\usepackage{mathrsfs}
\usepackage{amsfonts}
\usepackage{xspace}
\allowdisplaybreaks

\begin{document}
\title{The effect of the gravitational constant variation on the propagation of gravitational waves}

\author{Jiachen An} 
\affiliation{Institute for Frontiers in Astronomy and Astrophysics, Beijing Normal University, Beijing 102206, China}
\affiliation{Department of Astronomy, Beijing Normal University, Beijing 100875, China}
\author{Yadong Xue} 
\affiliation{Institute for Frontiers in Astronomy and Astrophysics, Beijing Normal University, Beijing 102206, China}
\affiliation{Department of Astronomy, Beijing Normal University, Beijing 100875, China}
\author{Zhoujian Cao
\footnote{corresponding author}} \email[Zhoujian Cao: ]{zjcao@amt.ac.cn}
\affiliation{Institute for Frontiers in Astronomy and Astrophysics, Beijing Normal University, Beijing 102206, China}
\affiliation{Department of Astronomy, Beijing Normal University, Beijing 100875, China}
\affiliation{School of Fundamental Physics and Mathematical Sciences, Hangzhou Institute for Advanced Study, UCAS, Hangzhou 310024, China}
\author{Xiaokai He} 
\affiliation{School of Mathematics and Statistics, Hunan First Normal University, Changsha 410205, China}
\author{Bing Sun} 
\affiliation{CAS Key Laboratory of Theoretical Physics, Institute of Theoretical Physics,
Chinese Academy of Sciences, Beijing 100190, China}

\begin{abstract}
Since the first detection of gravitational waves, they have been used to investigate various fundamental problems, including the variation of physical constants. Regarding the gravitational constant, previous works focused on the effect of the gravitational constant variation on the gravitational wave generation. In this paper, we investigate the effect of the gravitational constant variation on the gravitational wave propagation. The Maxwell-like equation that describes the propagation of gravitational waves is extended in this paper to account for situations where the gravitational constant varies. Based on this equation, we find that the amplitude of gravitational waves will be corrected. Consequently the estimated distance to the gravitational wave source without considering such a correction may be biased. Applying our correction result to the well known binary neutron star coalescence event GW170817, we get a constraint on the variation of the gravitational constant. Relating our result to the Yukawa deviation of gravity, we for the first time get the constraint of the Yukawa parameters in 10Mpc scale. This scale corresponds to a graviton mass $m_g\sim10^{-31}$eV.
\end{abstract}

\maketitle

\section{Introduction}
The fundamental constants of physics including the gravitational constant, the speed of light, Planck's constant, the charge and the mass of an electron provide us a set of absolute units for measurements of physical quantities \cite{liangcao2021}. More importantly these fundamental constants are a cornerstone of our physical laws \cite{RevModPhys.75.403,Uzan2011}. Especially the variation of any fundamental constant is closely related to a violation of the equivalence principle. Since the equivalence principle is the foundation of general relativity, the variation of fundamental constants is a good model independent test of general relativity. In the current paper we focus on the variation of the gravitational constant $G$. Theoretically many gravity theories beyond Newtonian gravity and general relativity can be casted phenomenologically as space and/or time dependence of the gravitational constant \cite{2002ApJ...570..463L,PhysRevD.97.104068,2003sttg.book.....F}.

The large numbers hypothesis proposed by Dirac implies that these fundamental constants may vary respect to time \cite{DIRAC1937,PhysRevD.91.121101}. Kinds of experiments and observations have been used to set the limit of the time variation of the gravitational constant $\left|\dot{G}/G\right|$ including the solar evolution ($\sim10^{-10}{\rm yr}^{-1}$) \cite{PhysRevLett.36.833}, lunar occultations and eclipses ($\sim10^{-11}{\rm yr}^{-1}$) \cite{1973Natur.241..519M}, paleontological evidences ($\sim10^{-11}{\rm yr}^{-1}$) \cite{PhysRevLett.72.454}, white dwarf cooling and pulsations ($\sim10^{-10}{\rm yr}^{-1}$) \cite{2013JCAP...06..032C}, neutron star masses and ages ($\sim10^{-12}{\rm yr}^{-1}$) \cite{PhysRevLett.77.1432}, star cluster evolutions ($\sim10^{-12}{\rm yr}^{-1}$) \cite{1996A&A...312..345D}, big bang nucleosynthesis abundances ($\sim10^{-12}{\rm yr}^{-1}$) \cite{2020EPJC...80..148A}, asteroseismology ($\sim10^{-12}{\rm yr}^{-1}$) \cite{2019ApJ...887L...1B}, lunar laser ranging ($\sim10^{-14}{\rm yr}^{-1}$) \cite{2018CQGra..35c5015H}, evolutions of planetary orbits ($\sim10^{-14}{\rm yr}^{-1}$)  \cite{2018NatCo...9..289G}, binary pulsars ($\sim10^{-12}{\rm yr}^{-1}$) \cite{2019MNRAS.482.3249Z}, high-resolution quasar spectra ($\sim10^{-14}{\rm yr}^{-1}$) \cite{2021GReGr..53...37L}, gravitational wave observations of binary neutron stars ($\sim10^{-8}{\rm yr}^{-1}$) \cite{PhysRevLett.126.141104,WANG2022137416} and supernovae \cite{2018JCAP...10..052Z}. The time variation of $G$ may be related to the Hubble tension \cite{PhysRevD.104.L021303}.

The position dependence of $G$ means the breaking down of the inverse-square law for gravity \cite{1999snng.book.....F}. The most plausible form of position dependence of $G$ is the Yukawa form \cite{1986PhLB..180..213D,1999snng.book.....F,2003ARNPS..53...77A,2009PrPNP..62..102A}
\begin{align}
G=G_\infty[1+\alpha(1+\frac{r}{\lambda}) e^{-\frac{r}{\lambda}}],
\end{align}
where $r$ is the distance to the gravity source, $\alpha$ and $\lambda$ are two parameters of the Yukawa model, and $G_\infty$ corresponds to the gravitational constant at infinity. Many laboratory experiments have been contributed to constrain the parameters $\alpha$ and $\lambda$ in submeter range \cite{2004PhRvD..70d2004H,PhysRevLett.124.051301,PhysRevLett.124.101101,2021EL....13511001L,PhysRevLett.126.211101,PhysRevD.104.L061101}. Geophysical observations, laser satellite ranging \cite{2002PhLA..298..315I}, laser lunar ranging \cite{2005gr.qc.....9114M,Merkowitz2010}, planetary orbits \cite{2014RAA....14..139L,PhysRevLett.123.161103} and S stars orbiting at the center of our galaxy \cite{2022arXiv221112951J} are used to constrain the parameters $\alpha$ and $\lambda$ within the galaxy scale. The parameter $\lambda$ can be related to Compton wavelength of a graviton \cite{2003ARNPS..53...77A,2016JCAP...05..045Z}. Using the velocity measurement of gravitational wave and the dissipation relation of gravitons, $\lambda$ can be constrained to $\lambda\gtrsim10^{15}$m ($m_g<10^{-23}$eV) by GW170817 \cite{2022arXiv220515432S}.

The gravitational constant $G$ will affect the mass of white dwarfs and neutron stars. The related gravitational wave to neutron stars and supernovae will be consequently affected. The time variation of $G$ can be determined by such gravitational waves \cite{PhysRevLett.126.141104,2018JCAP...10..052Z}. If the gravitational constant $G$ varies during the generating process of gravitational waves, the waveform of the gravitational wave will be affected by such variations \cite{Yunes2010,PhysRevD.98.084042,PhysRevD.100.104001}. The resulted waveform can be used as the template to analyze gravitational wave data. That's the traditional way to measure the variation of $G$ through gravitational wave detections \cite{WANG2022137416,2022arXiv220515432S,2023SCPMA..6620411G}.

If the gravitational constant $G$ varies during the propagating process of gravitational waves, one may suspect that the gravitational wave will be affected accordingly. To the best of our knowledge, there is no investigation of this problem yet. The current paper focuses on this problem. In the next section, we review the Maxwell-like equation for the gravitational wave propagation. After that, we assume such equation is still valid when the gravitational constant $G$ varies respect to space. In Sec.~\ref{sec3} we analyze the Maxwell-like equation and find that the amplitude of the gravitational wave will be corrected by the varying gravitational constant $G$. Then we apply our result to the events GW170817 to measure the variation of $G$ and the Yukawa parameters $\alpha$ and $\lambda$. Finally we summarize our paper in Sec.~\ref{sec4} with some discussions.
\section{Maxwell-like equations for gravitational wave propagations}
The propagation of gravitational waves can be described as a perturbation of a Minkowski spacetime
\begin{align}
g_{\mu\nu}=\eta_{\mu\nu}+h_{\mu\nu}.
\end{align}
Following the traditional notation we have the trace-inverse tensor
\begin{align}
\bar{h}_{\mu\nu}=h_{\mu\nu}-\frac{1}{2}h\eta_{\mu\nu},
\end{align}
where $h$ is the trace of $h_{\mu\nu}$. Within Lorenz gauge
\begin{align}
-\frac{1}{c}\frac{\partial}{\partial t}\bar{h}_{0\mu}+\frac{\partial}{\partial x^i}\bar{h}_{i\mu}=0\label{eq4}
\end{align}
we have the linearized Einstein equation
\begin{align}
-\frac{1}{c^2}\frac{\partial^2}{\partial t^2}\bar{h}_{\mu\nu}+\nabla^2\bar{h}_{\mu\nu}=-16\pi\frac{G}{c^4}T_{\mu\nu}.\label{eq1}
\end{align}

If we introduce electromagnetism-like notation \cite{wald84}
\begin{align}
&A_\mu\equiv-\frac{1}{4}c\bar{h}_{0\mu},\\
&J_\mu\equiv-\frac{T_{0\mu}}{c},
\end{align}
the linearized Einstein equation (\ref{eq1}) results in
\begin{align}
-\frac{1}{c^2}\frac{\partial^2}{\partial t^2}A_\mu+\nabla^2A_\mu=-4\pi\frac{G}{c^2}J_\mu.\label{eq2}
\end{align}
And the Lorenz gauge condition (\ref{eq4}) results in
\begin{align}
-\frac{1}{c}\frac{\partial}{\partial t}A_0+\frac{\partial}{\partial x^i}A_i=0.\label{eq5}
\end{align}

If we introduce more electromagnetism-like notation
\begin{align}
&\varphi\equiv cA_0,\\
&\vec{A}\equiv A_i,\\
&\vec{E}\equiv \nabla\varphi-\frac{\partial\vec{A}}{\partial t},\label{eq3}\\
&\vec{B}\equiv\nabla\times\vec{A}.\label{eq6}
\end{align}

The divergence of (\ref{eq3}) gives
\begin{align}
\nabla\cdot\vec{E}=\nabla^2\varphi-\frac{\partial\nabla\cdot\vec{A}}{\partial t}.\label{eq10}
\end{align}
Based on (\ref{eq5}) we have
\begin{align}
\nabla\cdot\vec{A}=\frac{1}{c^2}\frac{\partial}{\partial t}\varphi.
\end{align}
Consequently Eq.~(\ref{eq10}) becomes
\begin{align}
\nabla\cdot\vec{E}&=\nabla^2\varphi-\frac{1}{c^2}\frac{\partial^2\varphi}{\partial t^2}\\
&=c\left(\nabla^2A_0-\frac{1}{c^2}\frac{\partial^2}{\partial t^2}A_0\right)\\
&=-4\pi\frac{G}{c}J_0
\end{align}
where we have used Eq.~(\ref{eq2}) in the last step.

The curl of (\ref{eq3}) gives
\begin{align}
\nabla\times\vec{E}&=-\frac{\partial\nabla\times\vec{A}}{\partial t}\\
&=-\frac{\partial\vec{B}}{\partial t}
\end{align}
where we have used Eq.~(\ref{eq6}) in the last step.

Directly from Eq.~(\ref{eq6}) we have
\begin{align}
&\nabla\cdot\vec{B}=0.
\end{align}

Taking the curl of Eq.~(\ref{eq6}) we have
\begin{align}
\nabla\times\vec{B}&=\nabla\times(\nabla\times\vec{A})\\
&=\nabla(\nabla\cdot\vec{A})-\nabla^2\vec{A}\\
&=\nabla(\frac{1}{c^2}\frac{\partial}{\partial t}\varphi)+4\pi\frac{G}{c^2}\vec{j}-\frac{1}{c^2}\frac{\partial^2}{\partial t^2}\vec{A}\\
&=\frac{1}{c^2}\frac{\partial}{\partial t}\left[\nabla\varphi-\frac{\partial}{\partial t}\vec{A}\right]+4\pi\frac{G}{c^2}\vec{j}\\
&=\frac{1}{c^2}\frac{\partial\vec{E}}{\partial t}+4\pi\frac{G}{c^2}\vec{j}.
\end{align}

In summary we get Maxwell-like equations
\begin{align}
&\nabla\cdot(\epsilon\vec{E})=-\rho,\label{eq8}\\
&\nabla\cdot\vec{B}=0,\\
&\nabla\times\vec{E}=-\frac{\partial\vec{B}}{\partial t},\\
&\nabla\times\frac{\vec{B}}{\mu}=\vec{j}+\frac{\partial(\epsilon\vec{E})}{\partial t},\label{eq9}
\end{align}
where
\begin{align}
&\rho\equiv\frac{J_0}{c},\\
&\vec{j}\equiv J_i,\\
&\epsilon\equiv\frac{1}{4\pi G},\\
&\mu\equiv\frac{4\pi G}{c^2}.
\end{align}

We note that the above Maxwell-like equations can only determine $\bar{h}_{0\mu}$ while leave $\bar{h}_{ij}$ alone. In source free cases $\rho=0,\vec{j}=0$, if more $\bar{h}_{ij}=0$ the transverse-traceless gauge transformation will indicate that $\vec{E}$ and $\vec{B}$ are pure gauges. Generally $\vec{E}$ and $\vec{B}$ should be combined with $\bar{h}_{ij}$ to determine gravitational wave. Alternative Maxwell-like equation descriptions of gravitational waves can be found in Ref.~\cite{He2019}.

\section{The effect of the position dependent gravitational constant on the propagation of gravitational waves \label{sec3}}
In the following, we assume that the Maxwell-like equations \eqref{eq8}-\eqref{eq9} are still valid when the gravitational constant $G$ is not constant any more. In the current work, we consider a position-dependent $G$. This consideration is motivated by the possibility that, when strong equivalence principle is violated, $G$ may take different values at different gravitational environment.

We consider gravitational waves propagating in source free regions and introduce electromagnetic notation
\begin{align}
&\vec{D}\equiv\epsilon\vec{E},\\
&\vec{H}\equiv\frac{\vec{B}}{\mu}.
\end{align}
Then the above source free Maxwell-like equations take form
\begin{align}
&\nabla\cdot\vec{D}=0,\label{eq11}\\
&\nabla\cdot\vec{B}=0,\\
&\nabla\times\vec{E}=-\frac{\partial\vec{B}}{\partial t},\\
&\nabla\times\vec{H}=\frac{\partial\vec{D}}{\partial t},\label{eq12}
\end{align}
which describes the gravitational wave propagation in vacuum.

We put the gravitational wave source at the coordinate original point. The transverse property of gravitational waves makes sure
\begin{align}
&\vec{X}=X_\theta\hat{e}_\theta+X_\phi\hat{e}_\phi
\end{align}
in spherical coordinate system $(r,\theta,\phi)$, here $X$ represents any of $\vec{E},\vec{D},\vec{B},\vec{H}$.

Calculating Eqs.~(\ref{eq11})-(\ref{eq12}) straightforwardly we get
\begin{align}
    &\begin{cases}
        \frac{1}{r}\frac{\partial}{\partial r}(r{E}_\theta)&=-\mu\frac{\partial}{\partial t}{H}_\phi,\\
        \frac{1}{r}\frac{\partial}{\partial r}(r{E}_\phi)&=\mu\frac{\partial}{\partial t}{H}_\theta,\\
        \frac{1}{r}\frac{\partial}{\partial r}(r{H}_\theta)&=\epsilon\frac{\partial}{\partial t}{E}_\phi,\\
        \frac{1}{r}\frac{\partial}{\partial r}(r{H}_\phi)&=-\epsilon\frac{\partial}{\partial t}{E}_\theta
    \end{cases}\\
    &\begin{cases}
        \frac{\partial}{\partial r}(r{E}_\theta)&=-\mu\frac{\partial}{\partial t}(r{H}_\phi),\\
        \frac{\partial}{\partial r}(r{E}_\phi)&=\mu\frac{\partial}{\partial t}(r{H}_\theta),\\
        \frac{\partial}{\partial r}(r{H}_\theta)&=\epsilon\frac{\partial}{\partial t}(r{E}_\phi),\\
        \frac{\partial}{\partial r}(r{H}_\phi)&=-\epsilon\frac{\partial}{\partial t}(r{E}_\theta)
    \end{cases}\\
    &\begin{cases}
        \mu\frac{\partial}{\partial r}[\mu^{-1}\frac{\partial}{\partial r}(r{E}_\theta)]&=-\mu\frac{\partial}{\partial t}\frac{\partial}{\partial r}(r{H}_\phi)
        =\epsilon\mu\frac{\partial}{\partial t}\frac{\partial}{\partial t}(r{E}_\theta),\\
        \mu\frac{\partial}{\partial r}[\mu^{-1}\frac{\partial}{\partial r}(r{E}_\phi)]&=\mu\frac{\partial}{\partial t}\frac{\partial}{\partial r}(r{H}_\theta)
        =\epsilon\mu\frac{\partial}{\partial t}\frac{\partial}{\partial t}(r{E}_\phi),\\
        \epsilon\frac{\partial}{\partial r}[\epsilon^{-1}\frac{\partial}{\partial r}(r{H}_\theta)]&=\mu\frac{\partial}{\partial t}\frac{\partial}{\partial r}(r{E}_\phi)
        =\epsilon\mu\frac{\partial}{\partial t}\frac{\partial}{\partial t}(r{H}_\theta),\\
        \epsilon\frac{\partial}{\partial r}[\epsilon^{-1}\frac{\partial}{\partial r}(r{H}_\phi)]&=-\mu\frac{\partial}{\partial t}\frac{\partial}{\partial r}(r{E}_\theta)
        =\epsilon\mu\frac{\partial}{\partial t}\frac{\partial}{\partial t}(r{H}_\phi)
    \end{cases}
\end{align}
Equivalently we have
\begin{align}
\label{radiation_part}
    \begin{cases}
        \frac{\partial}{\partial r}\frac{\partial}{\partial r}(r{E}_\theta)-\frac{\partial}{\partial r}(\ln\mu)\frac{\partial}{\partial r}(r{E}_\theta)-\frac{\partial}{\partial (ct)}\frac{\partial}{\partial (ct)}(r{E}_\theta)=0,\\
        \frac{\partial}{\partial r}\frac{\partial}{\partial r}(r{E}_\phi)-\frac{\partial}{\partial r}(\ln\mu)\frac{\partial}{\partial r}(r{E}_\phi)-\frac{\partial}{\partial (ct)}\frac{\partial}{\partial (ct)}(r{E}_\phi)=0,\\
        \frac{\partial}{\partial r}\frac{\partial}{\partial r}(r{H}_\theta)-\frac{\partial}{\partial r}(\ln\epsilon)\frac{\partial}{\partial r}(r{H}_\theta)-\frac{\partial}{\partial (ct)}\frac{\partial}{\partial (ct)}(r{H}_\theta)=0,\\
        \frac{\partial}{\partial r}\frac{\partial}{\partial r}(r{H}_\phi)-\frac{\partial}{\partial r}(\ln\epsilon)\frac{\partial}{\partial r}(r{H}_\phi)-\frac{\partial}{\partial (ct)}\frac{\partial}{\partial (ct)}(r{H}_\phi)=0.
    \end{cases}
\end{align}

All equations in \eqref{radiation_part} have the form
\begin{equation}\label{radiation}
    \frac{\partial^2}{\partial r^2}f(r,t)-p(r)\frac{\partial}{\partial r}f(r,t)-\frac{\partial^2}{\partial (ct)^2}f(r,t)=0
\end{equation}
with $p$ corresponding to $\frac{\partial}{\partial r}(\ln\mu)$ or $\frac{\partial}{\partial r}(\ln\epsilon)$.

If $p$ is a constant, we can find the out-going wave solution
\begin{align}
f(r,t)=Ce^{(p/2)r}e^{i(\sqrt{(\omega/c)^2-(p/2)^2}r-\omega t)}
\end{align}

If $\frac{dp}{dr}$ is small, we can divide the domain in question into many spherical shells, and in each of them $p$ is nearly constant. Noting that the frequency $\omega$ remains constant during propagation, we have an approximated solution
\begin{align}
f(r,t)&=Ce^{\int(p/2)dr}e^{i(\int\sqrt{(\omega/c)^2-(p/2)^2}dr-\omega t)}\label{wave_solution}\\
    &\approx Ce^{\int(p/2)dr}e^{i(\frac{\omega D_L}{c}-\omega t)}.
\end{align}
So corresponding to \eqref{radiation_part} we have
\begin{align}
&r\vec{E} \propto e^{\int(\frac{d}{d r}(\ln\mu)/2)d r} \propto \mu^{1/2},\\
&r\vec{H}\propto e^{\int(\frac{d}{d r}(\ln\epsilon)/2)d r} \propto \epsilon^{1/2}.
\end{align}
Since $\vec{H}=\mu^{-1}\nabla\times\vec{A}$ and $\vec{H}\propto \epsilon^{1/2}/r$, we get
\begin{align}
\mu^{-1}(\omega/c)\vec{A}\propto\epsilon^{1/2}/r
\end{align}
which means
\begin{align}
\vec{A}\propto \mu^{1/2}/r.
\end{align}
Since $\vec{E}=-\vec{\nabla}\varphi-\frac{\partial}{\partial t}\vec{A}$, $\vec{E}\propto \mu^{1/2}/r$ and $\frac{\partial}{\partial t}\vec{A}\propto \omega\mu^{1/2}/r\propto \mu^{1/2}/r$, we get
\begin{align}
(\omega/c)\varphi\propto\mu^{1/2}/r
\end{align}
which means
\begin{align}
\varphi/c\propto\mu^{1/2}/r.
\end{align}
Therefore, we have
\begin{align}
\bar{h}_{0 \mu}=-4(\varphi/c,\vec{A})/c\propto\mu^{1/2}/cr\propto G^{1/2}/c^2r.
\end{align}

If the usual gravitational waveform without considering the variation of $G$ is known as $h_0\equiv h_{0+}-ih_{0\times}=Ae^{i\Phi}$, the above approximated solution means the corrected waveform is
\begin{align}
h=\sqrt\frac{G_d}{G_s}Ae^{i\Phi},
\end{align}
where $G_s$ means the gravitational constant at the gravitational wave source and $G_d$ is the gravitational constant at the detector. That is to say just the amplitude of the waveform is corrected by a factor $\sqrt\frac{G_d}{G_s}$ while leaving the phase unchanged. More specifically the dependence of $A$ and $\Phi$ on the gravitational constant only involves $G_s$.

In the usual analysis of gravitational wave data, the luminosity distance of the source is determined by the amplitude of the waveform. That is to say the luminosity distance measured by gravitational waves is in fact
\begin{align}
D_{L,GW}\equiv\sqrt\frac{G_s}{G_d} D_L,\label{eq7}
\end{align}
when $G$ varies respect to space.

\begin{figure}
\begin{tabular}{c}
\includegraphics[width=0.45\textwidth]{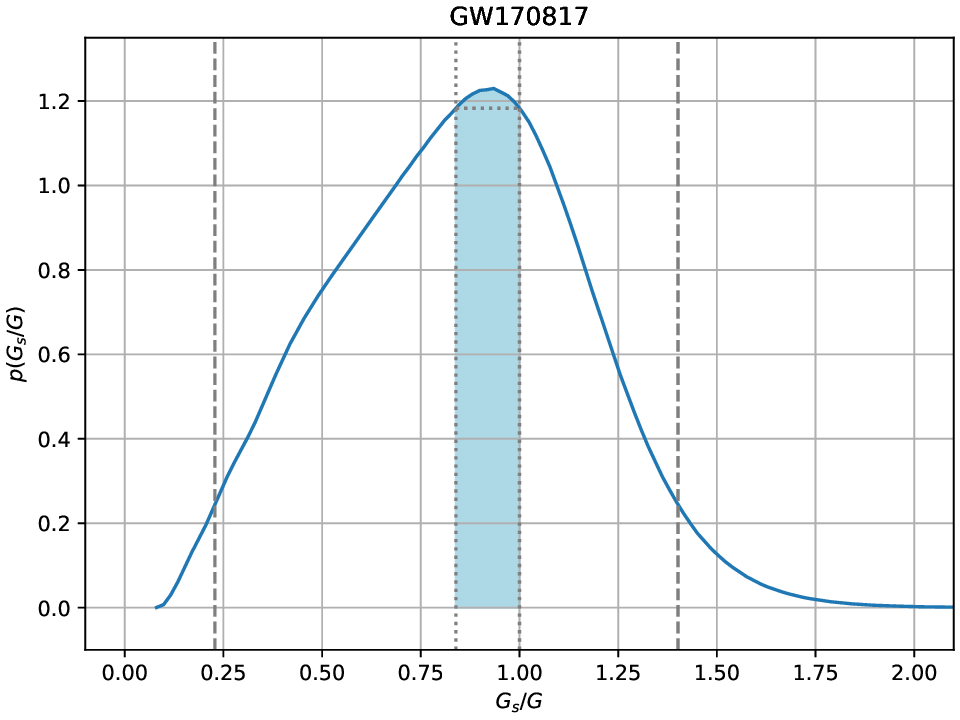}
\end{tabular}
\caption{Posterior distribution of $\frac{G_s}{G_d}$ according to GW170817/GRB170817A. The region between the two vertical dashed lines represents the parameter domain with 95\% confidence.}\label{fig1}
\end{figure}

The measurement results of GW170817 and GRB170817A give \cite{Abbott2017}
\begin{align}
D_{L,GW}&=40_{-14}^{+8}\text{Mpc}\\
D_L&=42.9_{-3.2}^{+3.2}\text{Mpc}.
\end{align}
And the corresponding parameter samples of posterior distribution have been released by LIGO-VIRGO collaboration \cite{gw-openscience,2021SoftX..1300658A}, Fermi Gamma-ray Burst Monitor, and INTEGRAL collaboration \cite{Abbott2017}. We calculate the probability distribution of $\frac{G_s}{G_d}$ with Monte Carlo method based on these posterior distribution samples. We separately draw samples for $D_{L,GW}$ and $D_{L}$. We draw the $D_{L,GW}$ samples from the posterior samples in version 3 data of GW170817 published on the Gravitational Wave Open Science Center \cite{gw-openscience,2021SoftX..1300658A}. This set of posterior samples were generated by using IMRPhenomPv2NRT model while assuming a low-spin prior. And we assume that the distribution of $D_{L}$ is Gaussian with mean value $42.9\text{Mpc}$ and variation $3.2\text{Mpc}$ provided by Ref.~\cite{Abbott2017}. We plot the result in Fig.~\ref{fig1}. $\frac{G_s}{G_d}=1$ corresponds to that $G$ is invariant. From the peak of the curve shown in Fig.~\ref{fig1} to the position $\frac{G_s}{G_d}=1$ corresponds to the area the data preferring varied $G$ (the shaded region in the figure). The probability left is about 80\%. That is to say the observation data of GW170817 and GRB170817A implies that $G$ is likely to be invariant within an 80\% confidence interval.

\begin{figure}
\begin{tabular}{c}
\includegraphics[width=0.5\textwidth]{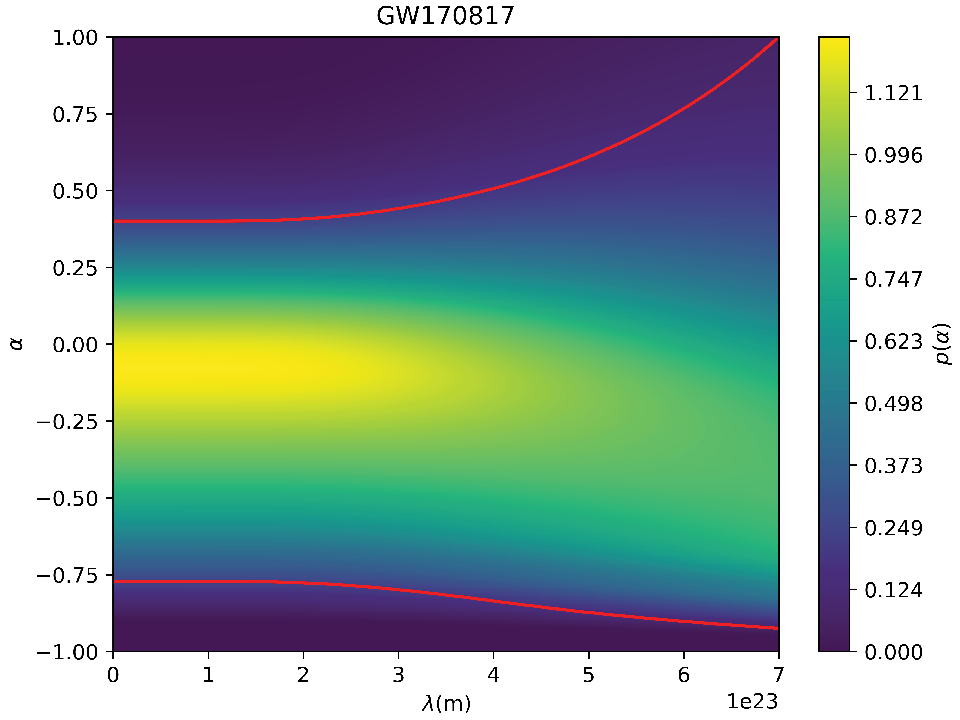}
\end{tabular}
\caption{Constraint of Yukawa parameters for $G$ according to GW170817/GRB170817A. The region between the two red lines represents the parameter domain with 95\% confidence.}\label{fig3}
\end{figure}

A Yukawa-type deviation from Newton's law of gravity gives a theoretical explanation of the position dependence of $G$ \cite{2003ARNPS..53...77A,2009PrPNP..62..102A}. In this Yukawa-type deviation the constraint \eqref{eq7} becomes
\begin{align}
\frac{1+\alpha}{1+\alpha(1+\frac{D_L}{\lambda})e^{-D_L/\lambda}}=\frac{G_s}{G_d}.
\end{align}
Using the samples for $D_{L,GW}$ and $D_{L}$ mentioned above, we can calculate the posterior distribution of $\frac{G_s}{G_d}$ which corresponds to the distribution of the left hand side of the above equation. For a given $\lambda$, such distribution reduces to posterior distributions of $\alpha$. We plot the posterior distributions of $\alpha$ respect to $\lambda$ in Fig.~\ref{fig3}. Compared to previous results, such as Fig.~10 of \cite{2009PrPNP..62..102A}, the constraint shown in Fig.~\ref{fig3} is the first result to be obtained in the region $\lambda\sim10^{23}$m $\sim10$Mpc, or $m_g\sim10^{-31}$eV. Previous results constraint $-0.1<\alpha<0.1$ for $10^{-2}$m $<\lambda<10^{15}$m. In the current work we get constraint $-0.75<\alpha<0.35$ for $10^{-2}$m$<\lambda<10^{23}$m which is consistent to the previous results on the one hand and extends the constraint scale to Mpc for the first time on the other hand.

\section{Conclusion and discussion}\label{sec4}
Starting from Dirac, many discussions have been given to the variation of fundamental constants, specifically the gravitational constant, which may vary from one place to another. If the equivalence principle breaks down, the gravitational constant may be different for places that admit different gravitational interaction strengths.

The variation of the gravitational constant will definitely affect the behavior of gravitational waves. There have been many works to investigate the generation of gravitational waves if the gravitational constant at the source varies respect to time. But to the best of our knowledge, the current work is the first one to study the effect of the variation of the gravitational constant on the propagation of gravitational waves.

In the current work we use a Maxwell-like equation to describe the propagation of gravitational waves. When the gravitational constant varies we assume such equation is still valid. Based on such propagation equation, we get the correction introduced by the variation of the gravitational constant. The correction changes the amplitude of the gravitational waves.

Due to the amplitude correction, the measured distance through gravitational waves will be biased. Through comparison of the biased distance measured by gravitational waves and the true one measured by electromagnetic waves, we can estimate the variation of the gravitational constant.

Based on the observation data of GW170817 and GRB170817A, we find that $G$ is likely to be invariant within an 80\% confidence interval. However, this conclusion assumes certain conditions and may be subject to further investigations. There are two possible ways deserving deeper study. One is looking for further away GW sources together with electromagnetic observations. The another way is searching for detail solutions to Eq.~(\ref{wave_solution}). When the high order phase corrections come in, more GW events and accurate phase detections can be used to constrain the variation of $G$.

If we assume more the variation of the gravitational constant can be described by a Yukawa-type deviation from Newton's law of gravity, we for the first time get the constraint of Yukawa parameter $\alpha$ at the scale 10Mpc.
\section*{Acknowledgments}
This research has made use of data or software obtained from the Gravitational Wave Open Science Center (gwosc.org), a service of LIGO Laboratory, the LIGO Scientific Collaboration, the Virgo Collaboration, and KAGRA. LIGO Laboratory and Advanced LIGO are funded by the United States National Science Foundation (NSF) as well as the Science and Technology Facilities Council (STFC) of the United Kingdom, the Max-Planck-Society (MPS), and the State of Niedersachsen/Germany for support of the construction of Advanced LIGO and construction and operation of the GEO600 detector. Additional support for Advanced LIGO was provided by the Australian Research Council. Virgo is funded, through the European Gravitational Observatory (EGO), by the French Centre National de Recherche Scientifique (CNRS), the Italian Istituto Nazionale di Fisica Nucleare (INFN) and the Dutch Nikhef, with contributions by institutions from Belgium, Germany, Greece, Hungary, Ireland, Japan, Monaco, Poland, Portugal, Spain. KAGRA is supported by Ministry of Education, Culture, Sports, Science and Technology (MEXT), Japan Society for the Promotion of Science (JSPS) in Japan; National Research Foundation (NRF) and Ministry of Science and ICT (MSIT) in Korea; Academia Sinica (AS) and National Science and Technology Council (NSTC) in Taiwan.

This work was supported in part by the National Key Research and Development Program of China Grant No. 2021YFC2203001 and in part by the NSFC (No.~11920101003, No.~12021003 and No.~12005016). Z. Cao was supported by ``the Interdiscipline Research Funds of Beijing Normal University" and CAS Project for Young Scientists in Basic Research YSBR-006.
\bibliography{refs}

\end{document}